\documentclass[a4paper]{aa}
\usepackage{psfig}
\usepackage{epsfig}
\usepackage{graphicx}

\def\ltsima{$\; \buildrel < \over \sim \;$}
\def\lsim{\lower.5ex\hbox{\ltsima}}
\def\gtsima{$\; \buildrel > \over \sim \;$}
\def\gsim{\lower.5ex\hbox{\gtsima}}

\def\src {SGR~1806--20}

\begin{document}


\title{INTEGRAL discovery of persistent hard X-ray emission from the
Soft Gamma-ray Repeater \src\thanks{Based on observations with
INTEGRAL, an ESA project with instruments and science data centre
funded by ESA member states (especially the PI countries: Denmark,
France, Germany, Italy, Switzerland, Spain), Czech Republic and
Poland, and with the participation of Russia and the USA. } }

\author{S. Mereghetti\inst{1}
        \and D. G\"{o}tz\inst{1}
        \and I.F. Mirabel\inst{2}
        \and K. Hurley\inst{3}}

\offprints{S. Mereghetti (sandro@mi.iasf.cnr.it)}

\institute{
        Istituto di Astrofisica Spaziale e Fisica Cosmica --
        Sezione di Milano ``G. Occhialini" -- INAF,
         via Bassini 15, I-20133 Milano, Italy
 \and
          European Southern Observatory,
          Alonso de Cordova 3107, Santiago, Chile
 \and
         University of California at Berkeley, Space Sciences Laboratory,
           Berkeley, CA 94720-7450, USA
 }

\date{Received  25 November 2004 ; Accepted: 16 February 2005 }

\authorrunning{S. Mereghetti et al.}

\titlerunning{{Discovery of hard X-ray emission from \src }}

\abstract{ We report the discovery of persistent hard X-ray
emission extending up to 150 keV from  the soft gamma-ray repeater
\src\ using data obtained with the INTEGRAL satellite in
2003-2004. Previous observations of hard X-rays from objects of
this class were limited to short duration bursts and rare
transient episodes of strongly enhanced luminosity (``flares'').
The emission observed with the IBIS instrument above 20 keV has a
power law spectrum with photon index in the range 1.5--1.9 and a
flux of 3 milliCrabs, corresponding to a 20-100 keV luminosity of
$\sim$10$^{36}$ erg s$^{-1}$ (for a distance of 15 kpc). The
spectral hardness and the luminosity correlate with the level of
source activity as measured from the number of emitted bursts.
  }

\maketitle

\section{Introduction}
\label{sect:intro}

Soft gamma-ray repeaters (SGRs) are a small group of high-energy
sources that were discovered  through  the observation of short
bursts in the hard X/soft $\gamma$-ray range (for a review see
Hurley 2000). The bursts have typical durations of the order of a
few hundred milliseconds and are emitted during sporadic
``active'' periods that can last from weeks to months. Their
spectra above $\sim$20 keV have been well described by optically
thin thermal bremsstrahlung with temperatures $\sim$30-40 keV (but
see Feroci et al. 2004 and Olive et al. 2004 for the spectra of
bursts at other energies).

For several years after their discovery,   SGRs could not be
unambiguously associated with persistent sources at other
wavelengths, with the exception of SGR 0525--66. The precise
localization of the latter source placed it in the Large
Magellanic Cloud supernova remnant N49, hinting at a neutron star
origin (Cline et al. 1982). More recent observations of the SGR
burst locations with X-ray imaging satellites led to the
identification of their persistent counterparts in the 0.5--10 keV
X-ray range. This proved to be crucial in confirming the neutron
star hypothesis through the discovery  of coherent pulsations
(periods of 7.5 and 5.2 s) and secular spin-down in the range
10$^{-11}$--10$^{-10}$ s s$^{-1}$ in two of these sources (\src\
and SGR 1900+14 ; Kouveliotou et al. 1998; Hurley et al. 1999a).
Furthermore, the association  with   clusters of massive stars
(Mirabel et al. 2000) indicates that SGRs are relatively young
objects. The most successful model advanced to explain the SGRs
involves neutron stars with a very high magnetic field, or
``magnetars'' (Duncan \& Thompson 1992, Thompson \& Duncan 1995).
In magnetars, which are assumed to have internal fields much
higher than the quantum critical value $B_{c} =
\frac{m_{e}^{2}c^{3}}{e\hbar}=4.4\times10^{13}$~G, the dominant
source of free energy is the magnetic field, rather than rotation
as in the ordinary radio pulsars. This energy is enough to power
both the bursts and the persistent X-ray emission
(L$\sim$10$^{34}$-10$^{35}$ erg s$^{-1}$).

The study of the persistent emission from SGRs has been  carried
out at energies below $\sim$10-20 keV to date. The only
observations of SGRs at higher energies, besides those of the
bursts, were limited to a small number of exceptionally bright
transient events, the so called ``giant'' and ``intermediate"
flares, which occurred in SGR 0525--66 (Mazets et al. 1979) and in
SGR 1900+14 (Hurley et al. 1999b, Guidorzi et al. 2004). Here we
report the discovery of emission extending up to 150 keV from the
quiescent counterpart of the soft gamma-ray repeater \src. Some
evidence for this high-energy component was reported in a
preliminary analysis of a subset of the data presented here
(G\"{o}tz et al. 2004b, Bird et al. 2004). In a companion paper
Molkov et al. (2005) report independent evidence from  different
INTEGRAL observations  of persistent hard X-ray ($>$20 keV)
emission from this source.

\begin{table*}[ht!]
\caption{{\it INTEGRAL Core Program Observations of \src\ }}
 \begin{center}
 \begin{tabular}{lccc}
 \hline
Observing       & Net exposure &  Flux 20-60 keV &  Flux 60-100 keV \\
Period           & (ksec)    &  (counts s$^{-1}$)   &  (counts s$^{-1}$)     \\
\hline
2003   March 12 - April 23        &  233  & 0.286$\pm$0.043    & 0.10$\pm$0.03 \\
2003  September 27 - October 15    & 278 &  0.344$\pm$0.041    & 0.09$\pm$0.03 \\
2004  February 17 - April 19      &  285 &  0.341$\pm$0.043    & 0.05$\pm$0.03 \\
2004  September 21 - October 14   &  213 &  0.511$\pm$0.050    & 0.19$\pm$0.03 \\
\hline
2003 March 12 - 2004 Oct. 14     &  1033 & 0.375$\pm$0.022      & 0.102$\pm$0.015 \\
\hline
\end{tabular}
\end{center}
\end{table*}

\begin{figure*}[th!]
\begin{center}
\psfig{figure=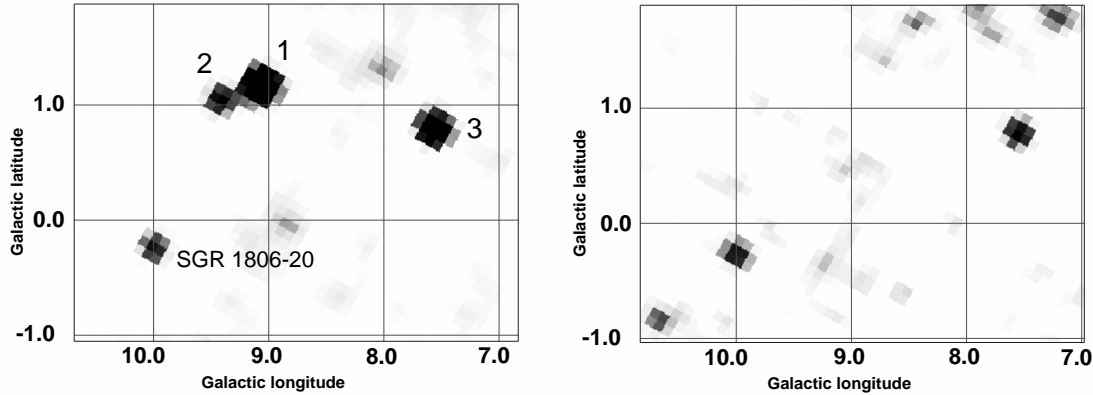,width=15cm}

\caption{Images of a  4$^{\circ}\times$3$^{\circ}$ region
containing the position of \src\ obtained with the IBIS/ISGRI
instrument in the 20-60 keV   (left) and  60-100 keV energy range
(right). The other detected sources are (1) GX 9+1, (2) IGR
J18027--2016 (Revnivtsev et al. 2004), and (3) IGR J17597--2201
(Lutovinov et al. 2003).}
\end{center}
\end{figure*}

\section{Data Analysis and Results}

 The location of \src\ has been repeatedly observed by the INTEGRAL
gamma-ray  satellite in 2003-2004 (Winkler et al. 2003) in the
course of Core Program (consisting of regular scans along the
Galactic plane and bulge), deep exposures of the Galactic Center
region, and Target of Opportunity observations. Here we
concentrate on the Core Program data obtained with the IBIS
instrument (Ubertini et al. 2003), which provides imaging above 15
keV with an unprecedented combination of sensitivity
($\sim$milliCrab) and angular resolution (12$'$) using the coded
mask imaging technique.

The available data  consist of more than 500 individual pointings
in which \src\ was observed at different off-axis angles. The
typical integration time of most pointings is $\sim$1800 s (those
obtained after May 2004 lasted $\sim$2200 s each). We restrict our
analysis to the pointings in which \src\ was in the central part
of the field of view (radius $<$8$^{\circ}$), where the
sensitivity is higher and the instrument calibration is better
known. This results in a total exposure time of one million
seconds (corrected for the vignetting effect). The observations,
spanning almost two years, were concentrated in four periods of
1-2 months each due to the visibility constraints of the
satellite, as indicated in Table~1.  We used the data obtained
with the IBIS lower energy detector ISGRI (Lebrun et al. 2003), an
array of 128$\times$128 CdTe pixels providing a geometric area of
2600 cm$^{2}$ in the nominal energy range 15 keV - 1 MeV.

The detector images in the 20-60 and 60-100 keV energy ranges of
the individual pointings were deconvolved, cleaned and co-added
using version 4.1 of the Offline Scientific Analysis software
(Goldwurm et al. 2003) provided by the INTEGRAL Science Data
Center (Courvoisier et al. 2003). A
4.5$^{\circ}\times$4.5$^{\circ}$ region of the resulting mosaics
is shown in Fig.~1. A source with a significance of 17.2$\sigma$
is detected in the lowest energy range at the Chandra position of
\src\ (Kaplan et al. 2002). Several other sources are visible in
the image (see figure caption for details). \src\ is also detected
in the 60-100 keV energy range, with a significance of
6.9$\sigma$. The source count rates of 0.375$\pm$0.022 counts
s$^{-1}$ and 0.102$\pm$0.015 counts s$^{-1}$ measured in the soft
and hard energy ranges correspond to a flux of $\sim$3 mCrabs.

Except for the period March-April 2003, the source was in a
bursting state during the reported observations, leading to the
detection of several bursts in the IBIS data\footnote{the analysis
of those observed in September-October 2003, showing an
anticorrelation between hardness and intensity of the bursts, has
been reported in G\"{o}tz et al. (2004a)}.
A few pointings
containing some strong bursts (Mereghetti et al. 2004a, Golenetski
et al. 2004) were removed from the analysis described above. The
effect of the remaining ones  is negligible: their cumulative
contribution of $\sim$16,000 counts in about 50 bursts,
corresponds to 4\% of the total number of source counts in the
integrated images.

Analyzing the data for the four observing periods separately in
the same way, we derived the fluxes reported in Table 1, which
give some evidence for an increase in the source hard X-ray
luminosity with time (see Fig.~3b). A fit of the 20-60 keV count
rate values with a constant flux gives a $\chi^2$=12.3,
corresponding to a 6.5$\times$10$^{-3}$ probability.

For the spectral analysis we produced images in 9 energy intervals
in the range 15-300 keV. The count rates extracted from the images
were fit using the latest available response matrices (RMF v11,
ARF v5). Since the first three observing periods yielded
consistent spectral parameters, we analyzed them together,
obtaining a best fit power law photon index $\Gamma$=1.9$\pm$0.2
and a 20-100 keV flux F$_{20-100}$=(4.7$\pm$0.5)$\times$10$^{-11}$
erg cm$^{-2}$ s$^{-1}$ (all errors are 1$\sigma$). During the last
observing campaign (Sept-Oct 2004) the corresponding  parameters
were $\Gamma$=1.5$\pm$0.3 and
F$_{20-100}$=(8.0$\pm$0.9)$\times$10$^{-11}$ erg cm$^{-2}$
s$^{-1}$. Thus there is some evidence for a harder spectrum
coupled with a flux increase. The two spectra are shown in Fig.~2a
and 2b. Note that the higher flux and harder spectrum of the
Sept.-Oct. 2004 period lead to a significant detection of \src\ up
to the 100-150 keV energy bin. For comparison, Fig.~2c shows the
typical spectrum of one of the bursts (note that the intensity has
been scaled down by a factor 10$^4$). It can be seen that the
bursts have a much softer spectrum than the persistent emission.

\section{Discussion}

Observations of \src\  below 10 keV carried out with  BeppoSAX and
ASCA indicated a power law spectrum with photon index $\sim$2 and
unabsorbed 2-10 keV flux of $\sim$1.6$\times$10$^{-11}$ erg
cm$^{-2}$ s$^{-1}$ (Mereghetti et al. 2000). Although the
detection of pulsations in the 10-20 keV range with RossiXTE
(G\"{o}\v{g}\"{u}\c{s} et al. 2002) showed that the emission from
\src\ extends above the soft X-ray range, no spectral information
was available above 10 keV and the BeppoSAX spectrum could be
fitted equally well by a thermal bremsstrahlung model with
temperature kT=11 keV. The extrapolation of the BeppoSAX power law
spectrum lies below the average value obtained with INTEGRAL. On
the other hand, a good  agreement is found between  the INTEGRAL
spectra and those measured below 10 keV  with  XMM-Newton   in
2003-2004 (Mereghetti et al. 2004b). This is shown by the dashed
lines in Fig. 2, which indicate the extrapolations of the
XMM-Newton power law spectra measured at dates consistent with the
INTEGRAL observations (Fig.2a: 7 October 2003; Fig.2b: 6 October
2004).

In Fig.~3 we compare the hard X-ray flux measured with INTEGRAL
with the level of activity from \src\ as a function of time. This
is indicated by the number of bursts detected by the
Interplanetary Network (IPN). The source was quiescent during the
first INTEGRAL observing period and moderately active in the two
following periods. The last INTEGRAL observations, in which a
harder and brighter persistent emission was detected, took place
after the strong reactivation of the Summer 2004. \src\ was still
particularly active in September-October 2004 (Mereghetti et al.
2004a, Golenetski et al. 2004).

The detection of pulsed hard X-rays (20-150 keV) from the
Anomalous X-ray Pulsar 1E 1841--045 has been recently reported by
Kuiper et al. (2004). Anomalous X-ray Pulsars (AXPs) share many
similarities with the SGRs (Mereghetti et al. 2002) and are also
thought to be magnetars (see Woods \& Thompson 2004 for a recent
review). It is therefore interesting to compare our results with
those obtained for this source. The high-energy emission from
1E~1841--045, with a power law photon index
$\Gamma$=1.47$\pm$0.05,  is definitely harder than that displayed
by  \src\ before September 2004. This value of the photon index,
obtained with the RXTE/HEXTE instrument, refers to the total
(pulsed plus unpulsed) flux and is consistent with the results of
the INTEGRAL flux measurements in the 18-60  and 60-120 keV ranges
(Molkov et al. 2004). The spectrum of the pulsed component in
1E~1841--045 is even harder ($\Gamma$=0.94$\pm$0.16; Kuiper et al.
2004). When compared to lower energy measurements, these results
imply that the spectrum of the AXP has a significant hardening
above $\sim$10 keV. This contrasts with the situation seen in
Figs.~2(a) and 2(b) for \src.

The hard X-ray emission from magnetars has been discussed in the
context of a model involving currents in a globally twisted
magnetosphere by Thompson, Lyutikov \& Kulkarni (2002). A
non-thermal tail in the X-ray spectrum is predicted as a result of
multiple resonant cyclotron scattering. According to these
authors, the differences between SGRs and AXPs, i.e.  the emission
of bursts in SGRs, as well as  their harder spectra and faster
spin down rate, are explained by a larger degree of twist in the
external magnetic field. Our observation of a harder persistent
spectrum when the bursting activity of \src\ was at its highest
level indicates that this correlation, previously seen when
comparing different sources, also holds within the same object.

\begin{figure}[t]
\psfig{figure=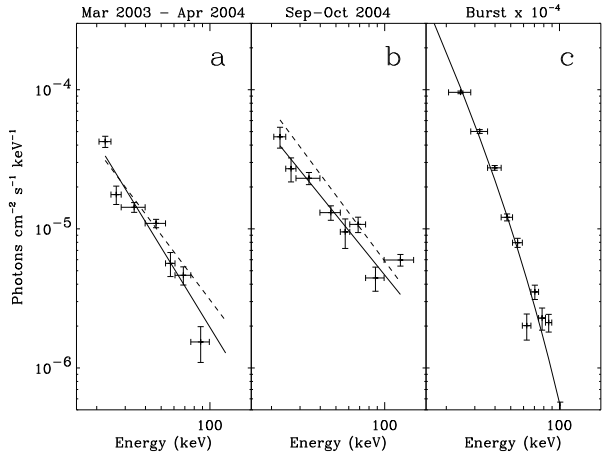,width=6cm,angle=90}
 \caption{IBIS/ISGRI spectra of \src: (a) persistent emission
 March 2003-Apr. 2004, (b) persistent emission
Sept.-Oct. 2004, (c) one burst (scaled down by a factor 10$^{4}$).
The solid lines are the best fits (power laws in (a) and (b),
thermal bremsstrahlung in (c)). The dashed lines indicate the
extrapolation of  power-law  spectra measured in the 1-10 keV
range with XMM-Newton (Mereghetti et al. 2004b).}
\end{figure}

\begin{figure}[t]
\psfig{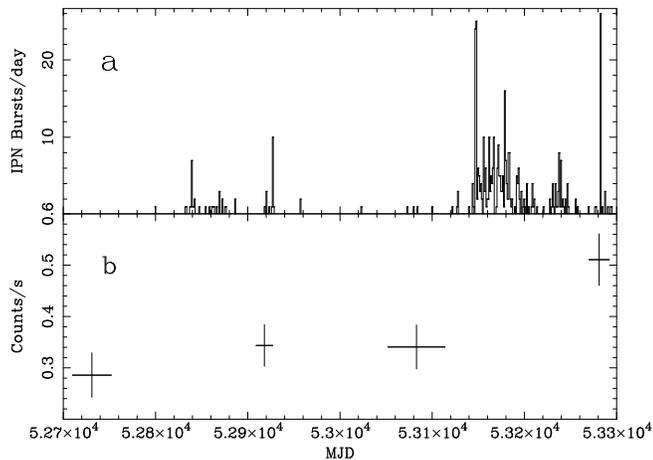}

\caption{(a) histogram of number of bursts per day detected by the
third Interplanetary Network (IPN). This histogram includes mostly
those bursts with fluences larger than 10$^{-7}$ erg cm$^{-2}$ and
is not corrected for experiment duty cycles and source visibility.
It therefore represents only the trend of the burst rate and not
its true value. (b) count rate in the 20-60 keV range. The
horizontal error bars indicate the time spanned by the (non
continuous) observations.}
\end{figure}

\section{Conclusions}

Thanks to the high sensitivity of the INTEGRAL IBIS instrument we
have discovered a hard X-ray component extending to 150 keV in the
persistent emission from \src. The imaging capabilities of IBIS
have been crucial for the observation of this crowded region of
the Galactic plane: they allow us to unambiguously associate the
observed hard X-ray emission with \src.

 The hard X-rays
correlate in intensity and spectral hardness with the level of
bursting activity, and, contrary to what is observed in the AXP
1E~1841--045, a comparison with lower energy data does not
indicate evidence for a spectral hardening at $\sim$10-20 keV. The
INTEGRAL data provide the first detection of persistent emission
in this energy range for a SGR and open a new important diagnostic
to study the physics of magnetars (see, e.g., Thompson \&
Beloborodov 2004).

\begin{acknowledgements}

This work has been partially supported by the Italian Space
Agency.  KH is grateful for support under NASA's INTEGRAL U.S.
Guest Investigator program, grant NAG5-13738.

\end{acknowledgements}

\end{document}